\documentclass[multphys,vecphys]{svmult}
\usepackage{makeidx}         
\usepackage{graphicx}        
\usepackage{multicol}        

\def\PZapgt{\ {\raise-.5ex\hbox{$\buildrel>\over\sim$}}\ }
\def\aplt{\ {\raise-.5ex\hbox{$\buildrel<\over\sim$}}\ }

\newcommand{\msun}{\mbox{${\rm M}_\odot$}}
\newcommand{\trlx}{\mbox{${t_{\rm rlx}}$}}
\newcommand{\rcore}{\mbox{${r_{\rm core}}$}}
\newcommand{\rhm}{\mbox{${r_{\rm hm}}$}}

\begin{document}

\title*{Reconstructing the Initial Relaxation Time of Young Star
Clusters in the Large Magellanic Cloud} 
\titlerunning{The Evolution of Star Clusters}

\author{Simon F. Portegies Zwart\inst{1} and Hui-Chen Chen\inst{1,2}}
\institute{$^1$Astronomical Institute 'Anton Pannekoek' and Section
Computational Science, University of Amsterdam, Kruislaan 403, 1098SJ
Amsterdam, the Netherlands \\ 
$^2$ Graduate Institute of Astronomy,
National Central University, No 300 Jhongda Rd. Jhongli City, Taiwan}

\maketitle

\begin{abstract}

We reconstruct the initial two-body relaxation time at the half mass
radius for a sample of young $\aplt 300$\,Myr star clusters in the
large Magellanic cloud. We achieve this by simulating star clusters
with 12288 to 131072 stars using direct $N$-body integration. The
equations of motion of all stars are calculated with high precision
direct $N$-body simulations which include the effects of the evolution
of single stars and binaries.

We find that the initial relaxation times of the sample of observed
clusters in the large Magellanic cloud ranges from about 200\,Myr to
about 2\,Gyr.  The reconstructed initial half-mass relaxation times
for these clusters has a much narrower distribution than the currently
observed distribution, which ranges over more than two orders of
magnitude.

\end{abstract}

\section{Introduction}
\label{Sec:Introduction}

The early evolution of young clusters of stars is of considerable
interest because of the rich observational data which became
available recently.  Studying young star clusters is also important
in order to obtain a better understanding of the conditions under
which clusters are born. The initial conditions of star
clusters have been debated actively over the years, but no consensus
has been reached either by observing or simulating the formation
process of stellar conglomerates. The main parameters which
characterize a star cluster at birth (and any time later) are the
number of stars, the mass function and the concentration of the
stellar distribution.

Part of the problem in determining the clusters' initial conditions
stems from our static view of the universe, our inability to run
simulations backward with time and our lack of understanding of the
physics of the star(cluster) formation process.  In this paper we
approach this problem by starting with a series of simulations in
which we fix most of the initial conditions but relax the initial
relaxation time. We subsequently calculate the evolution of the star
cluster with time. With this series of controlled numerical
experiments we trace back the observed parameters for a number of
young star clusters in the large Magellanic cloud along parallel
trajectories in parameter space.

\section{Simulations}

We focus on young $t \aplt 300$\,Myr star clusters, because excellent
observational data is available for a number of these
\cite{2003MNRAS.338...85M}.  For the simulations we adopt the Starlab
software environment \cite{2001MNRAS.321..199P}, which acquires it's
greatest speed with GRAPE-6 special purpose hardware
\cite{1997ApJ...480..432M,2003PASJ...55.1163M}\footnote{see {\tt
http://www.astrogrape.org}}. Our simulations are performed on the
GRAPE-6 hardware at the University of Tokyo and the
MoDeStA\footnote{see {\tt http://modesta.science.uva.nl}} platform at
the University of Amsterdam.

The simulated star clusters are initialized by selecting the number of
stars, stellar mass function, the density profile, binary fraction and
their orbital elements. For our most concentrated model (simulation
\#1) we adopt the initial conditions derived by Portegies Zwart et
al. \cite{2004Natur.428..724P} to mimic the 7-12\,Myr old star cluster
MGG-11 in the star-burst galaxy M82. In this paper, however, we extend
the evolution of this simulated cluster to about 100\,Myr
\cite{2006astro.ph..7461P}. Subsequent simulations are performed with
a larger cluster radius, resulting in a longer initial relaxation time
(for details on the simulation models \#\{1..4\}c with 128k stars as
listed in Tab.\,\ref{Tab:initials} and see
\cite{2006astro.ph..7461P}).  The stellar evolution model adopted is
based on \cite{1989ApJ...347..998E}, and the binaries are evolved with
SeBa \cite{1996A&A...309..179P}.

We summarize the selection of the initial conditions for simulation
\#1c (see Tab.\,\ref{Tab:initials}, see also
\cite{2006astro.ph..7461P}): first we selected 131072 stars
distributed in a King \cite{1966AJ.....71...64K} density profile with
$W_0=12$ and with masses from a Salpeter initial mass function ($x =
-2.35$) between 1\,$M_\odot$\, and 100\,$M_\odot$.  The total mass of
the cluster is then $M \simeq 433000\,M_\odot$.  The location in the
cluster where the stars are born is not correlated with the stellar
mass, i.e. there is no primordial mass segregation.  Ten percent of
the stars were randomly selected and provided with a companion
(secondary) star with a mass between 1\,$M_\odot$\, and the mass of
the selected (primary) star from a flat distribution.  The binary
parameters were selected as follows: first we chose a random binding
energy between $E = 10$\,kT (corresponding to a maximum separation of
about 1000\,$R_\odot$).  The maximum binding energy was selected such
that the distance at pericenter exceeded four times the radius of the
primary star. At the same time we select an orbital eccentricity from
the thermal distribution. If the distance between the stars at
pericenter is smaller than the sum of the stellar radii we select a
new semi-major axis and eccentricity. If necessary, we repeat this
step until the binary remains detached. As a result, binaries with
short orbital periods are generally less eccentric.  We ignored an
external tidal field of the Galaxy, but stars are removed from the
simulation if they are more than 60 initial half-mass radii ($\rhm$)
away from the density center of the cluster (100\,\rhm\, for the 12k
models, see Tab.\,\ref{Tab:initials}).

For the other simulations \#2c, \#3c and \#4c, we adopt the same
realization of the initial stellar masses, position and velocities (in
virial N-body units \cite{HM1986}) but with a different size and time
scaling to the stellar evolution, such that the two-body relaxation
time (${t_{\rm rlx}}$) for simulation \#2c is four times that of \#1c,
for simulation \#3c we used four times the two-body relaxation time of
what was used for simulation \#2c, etc. the initial conditions are
summarized in Tab.\,\ref{Tab:initials}.

We subsequently generate additional initial realizations for clusters
with a smaller number of stars (12k stars for models \#\{1..4\}a
versus 128k for models \#\{1..4\}c). To study the effect of the
initial density profile on the results are perform another set of
simulations with 12k stars but with a $W_0=6$ initial density profile
rather than the highly concentrated $W_0=12$.  The simulations with
$N=12$k are constructed without primordial binaries. 

\begin{table*}
\centering
\caption[]{Conditions for the performed calculations with a range of
number of stars (1k~$\equiv$~1024 stars) and cluster virial radii. The
columns give the model name, the number of stars in the initial model,
the concentration parameter ($W_0$), the initial half-mass relaxation
time ($t^i_{\rm rlx}$), the initial virial radius and core radius and
finally the initial crossing time.
%
}
\bigskip
\renewcommand{\arraystretch}{1.4}
\begin{tabular}  {lccccccc}
\hline
Run  & $N$ & $W_0$ &$t^i_{\rm rlx}$& $r_{\rm vir}$ & $r_{\rm core}$ & $t_{\rm ch}$  \\ 
     &     &       &  [Myr]     & [pc]          &  [pc]          & [Myr]         \\
\hline
\#1a & 12k & 12~~~ &   73& 2.27~~~ & 0.071~~~ &  0.258~~~\\
\#1b & 12k &  6~~~ &   81& 2.77~~~ & 0.670~~~ &  0.352~~~\\
\#1c & 128k& 12~~~ &   80& 1.27~~~ & 0.010~~~ &  0.032~~~\\ 
\#2a & 12k & 12~~~ &  304& 5.72~~~ & 0.072~~~ &  1.05~~~\\
\#2b & 12k &  6~~~ &  310& 6.98~~~ & 1.93~~~ &  1.010~~~\\
\#2c & 128k& 12~~~ &  320& 3.20~~~ & 0.026~~~ &  0.129~~~\\
\#3a & 12k & 12~~~ & 1200&14.6~~~~ & 0.151~~~ &  4.17~~~~\\
\#3b & 12k &  6~~~ & 1280&17.8~~~~ & 4.28~~~~ &  4.19~~~~\\
\#3c & 128k& 12~~~ & 1300&  8.1~~~ & 0.066~~~ &  0.516~~~\\
\#4a & 12k & 12~~~ & 4990&36.3~~~~ & 0.317~~~ & 16.6~~~~\\
\#4b & 12k &  6~~~ & 4920&44.2~~~~ &14.9~~~~ & 16.5~~~~\\
\#4c & 128k& 12~~~ & 5100& 20.0 ~~~  & 0.162~~~ &  2.07~~~\\
\hline 
\label{Tab:initials}
\label{Tab:results}
\end{tabular}
\end{table*}

After initialization we synchronously calculate the evolution of the
stars and binaries, and solve the equations of motion for the stars in
the cluster. The calculations are continued to an age of about
100\,Myr.

\section{Results}

The main difference between simulations are the number of stars and
since we ignore the external tidal field the number of stars drops
only slightly during the simulations, but the half-mass relaxation
time increasing substantially from its initial value.  At the same
time the cluster structure changes by becoming less concentrated. The
latter is mainly attributed to mass segregation and stellar mass loss
(see also
\cite{1990ApJ...351..121C,1995MNRAS.276..206F,1998A&A...337..363P,1998ApJ...503L..49T,2003MNRAS.340..227B}).
In \cite{2006astro.ph..7461P} we compared some of the characteristics
of a subset of the here presented simulations with the observed sample
of young star clusters in the LMC.  Here, in this proceedings paper,
we limit ourselves in comparing the currently observed two-body
relaxation time and compare these with the results of our simulations.

\begin{figure}
\centering
   \includegraphics[height=10cm,angle=+90]{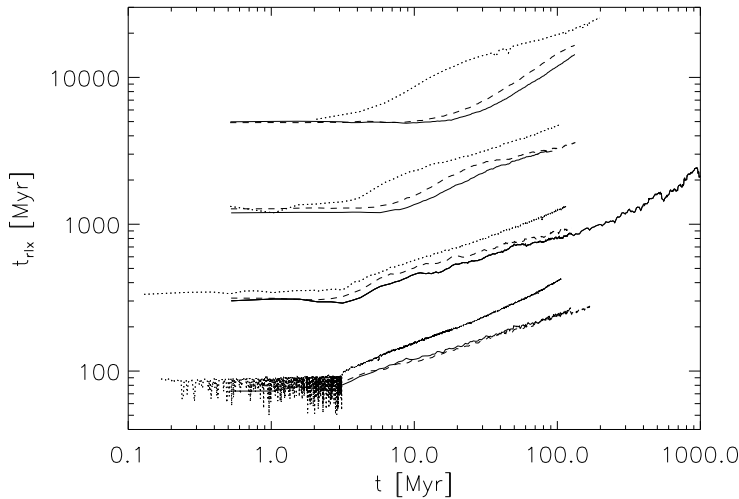}
   \caption{Evolution of the two-body relaxation time at the half mass
            radius for the four clusters \#1 (bottom), \#2, \#3 and
            \#4 (top).  The solid curves represent the data from
            models \#\{1..4\}a, dashes give \#\{1..4\}b and dotted
            curves gives the results for models \#\{1..4\}c.  Model
            \#2a was extended to an age of 1\,Gyr to demonstrate that
            the generally behavior doesn't suddenly change drastically.
	    \label{Fig:tvsTrlx_simulation} 
            }
\end{figure}

In figure\,\ref{Fig:tvsTrlx_simulation} we present the evolution of
the two-body relaxation time at the half-mass radius for the simulated
clusters \#1\{a..c\} to \#4\{a..c\}. The relaxation time is roughly
constant for the first $\sim 3$\,Myr for simulations \#1\{a..c\},
\#2\{a..c\}, \#3c and \#4c, to increase at later time.  For
simulations \#3a and \#3b the half-mass relaxation time starts to
increase only after about 6\,Myr, and after about 16\,Myr for
simulation \#4a and \#4b. We attribute this effect to the relatively
slow response of the latter models, in particular \#4a and \#4b, to
stellar mass loss. For these models mass loss is not adiabatic as is
the case for simulations \#1 and \#2, but rather impulsive (see
\cite{2006astro.ph..9061P} for a discussion in relation to variations
in the orbital parameters of binary star clusters).  The later
increase in the half-mass relaxation time is mainly caused by stellar
mass loss and, in a lesser extend by the internal structural changes
resulting from the internal dynamical evolution. The relaxation time
for the various simulation models tend to evolve in almost parallel
trajectories, with small variations.

\begin{figure}
\centering
   \includegraphics[height=10cm,angle=+90]{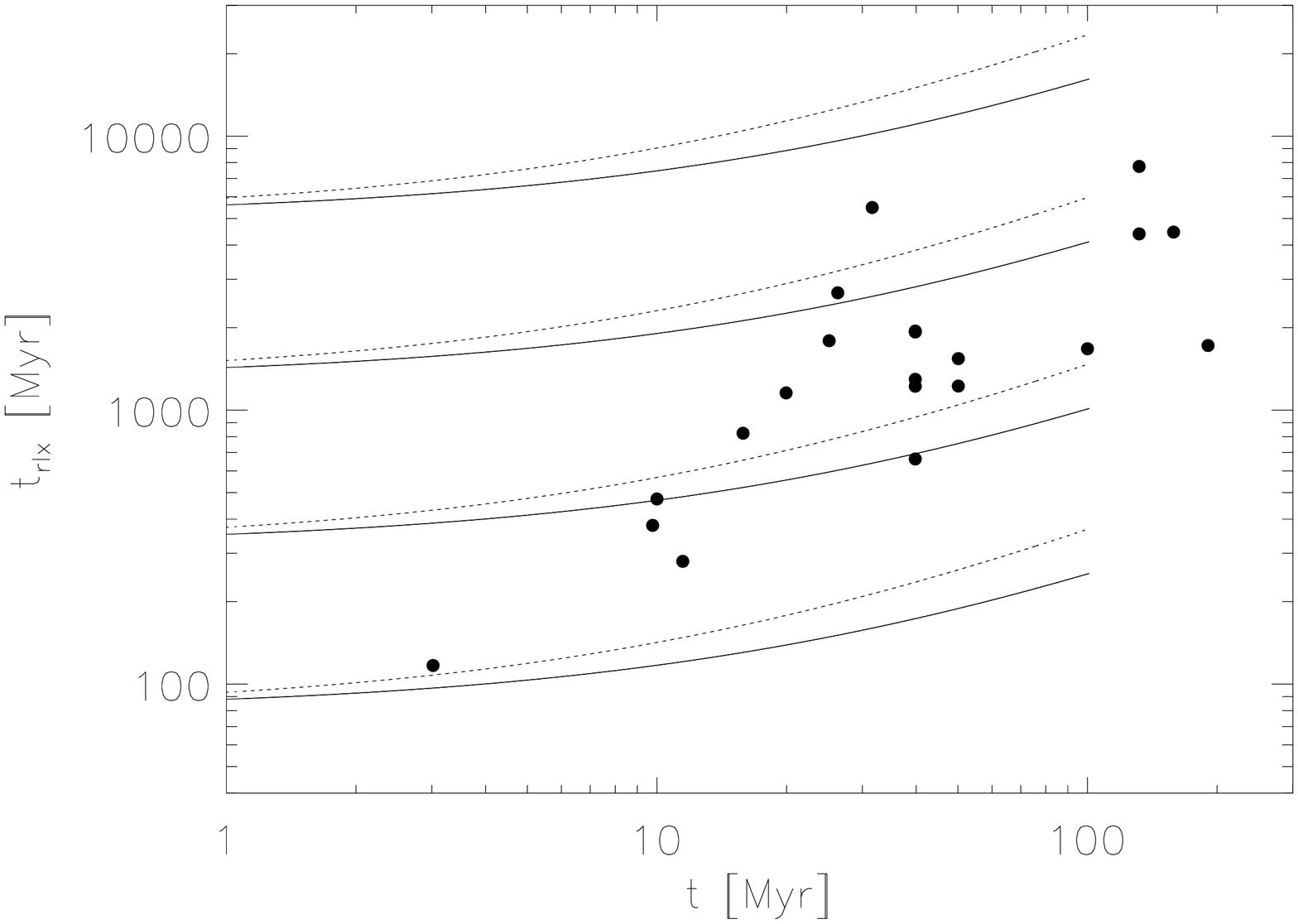}
   \caption{Approximated evolution of the two-body relaxation time at
            the half mass radius for the four clusters \#1 (bottom),
            \#2, \#3 and \#4 (top).  The solid curves represent the
            data from models with 12k stars
            (Eq.\,\ref{Eq:tvsTrlx_model}) and dotted curves gives the
            results for models 128k stars.  
            The bullets give an estimate for {\trlx} from the observed
            clusters \cite{2003MNRAS.338...85M} by adopting a mean
            mass of 0.5\,\msun\, and $\rhm = 3.92\rcore$.
  \label{Fig:tvsTrlx_model} } 
\end{figure}

We approximate the time evolution of the relaxation time with the
following two equations, for the $N=12$k  clusters we adopted
\begin{equation}
   \trlx(t) = \left( \kappa t^{2/3} + 1 \right) t^i_{\rm rlx}.
\label{Eq:tvsTrlx_model}
\end{equation}
Here $\kappa = 1/10$ for the 12k models and $\kappa = 1/6$ for the 128k
models.  The resulting tracks are presented in
Fig.\,\ref{Fig:tvsTrlx_model}.  

\begin{figure}
\centering
   \includegraphics[height=10cm,angle=+90]{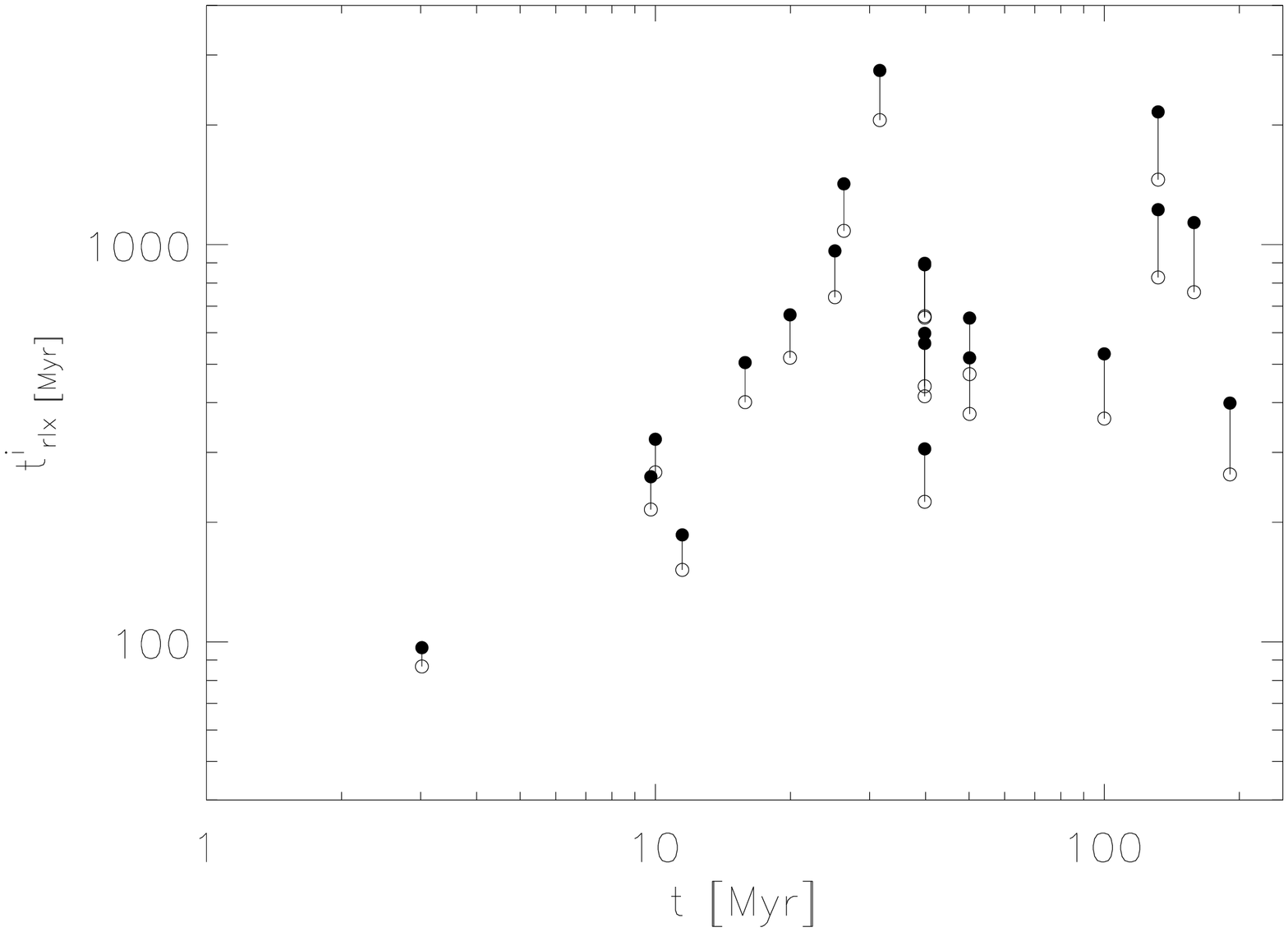}
   \caption{The initial half-mass relaxation time ($t^i_{\rm rlx}$) as
            a function of the measured age of the observed LMC
            cluster.  For each cluster we plot two symbols connected
            with a vertical line.  The lower (open circle) symbol
            indicates extrapolation with time using
            Eq.\,\ref{Eq:tvsTrlx_model} for 128k whereas the upper
            symbols (bullet) gives the result from inverting
            Eq.\,\ref{Eq:tvsTrlx_model} for 12k.
  \label{Fig:tvstrlx_initial} } 
\end{figure}

Since the evolution of the half-mass relaxation time for each set of
simulations with the same number of stars behaves quite similar, with
an initial offset, we decided to invert Eq.\,\ref{Eq:tvsTrlx_model} to
enable an extrapolation of the initial relaxation time for the
observed clusters. The result for the 21 observed clusters with an age
$\aplt300$\,Myr is presented in Fig.\,\ref{Fig:tvstrlx_initial}.  Here
we plot our estimate for the initial half-mass relaxation time
($t^i_{\rm rlx}$) as a function of the measured age of the observed
cluster. We ignore the cluster mass in the reconstruction of the
initial relaxation times for the observed clusters. From
Fig.\,\ref{Fig:tvsTrlx_simulation}, however, it may be clear that the
effect of the number of stars in the cluster is substantial. We
therefore opted for having the $N=12$k models to provide an estimate
of the upper limit and $N=128$k for providing the lower limit to our
estimate of the initial relaxation time. Those lower and upper limits
are presented in Fig.\,\ref{Fig:tvstrlx_initial} as open and filled
circles, respectively.

\section{Discussion and conclusions}

We performed extensive simulations of the young star clusters in the
large Magellanic cloud. The simulations ignore the external tidal
field, and we limited the initial mass function to a Salpeter between
1\,\msun\, and 100\,\msun.  We realize that the adopted initial mass
function may not be representative for the large Magellanic cloud, and
we performed several additional simulations using the initial mass
function proposed by \cite{2005imf..conf...77D}, which extends down to
the helium burning limit. The difference with the results presented
here, however, are quite small, but we tend to underestimate the
initial relaxation time compared to using a more realistic initial
mass function.

The half-mass relaxation time for the simulated clusters evolves on
almost parallel trajectories with an initial offset based on the
initially selected relaxation time. The trends in the relaxation time
is somewhat different from the smaller (in $N$) clusters than for the
larger clusters, as is evident in Fig.\,\ref{Fig:tvsTrlx_simulation}.
We attribute this divergence to the difference in the the initial
number of stars in these models, in particular since all other
parameters were kept as much as possible identical.  We note, however,
that our simulation with 128k stars were computed with 10\% primordial
hard binaries, whereas the smaller simulations with 12k stars were
computed without primordial binaries. This difference in the initial
conditions may attribute to the differences in the evolution of the
half-mass relaxation time. To test this hypothesis we performed
several additional simulations with 64k stars and no initial
binaries. Interestingly, the evolutionary tracks for the half-mass
relaxation time for these models is roughly situated between the 12k
and the 128k models. Based on these data we argue that the differences
we observe between the 12k and the 128k models can be attributed to
the differences in the initial number of stars, and that the presence
of primordial binaries has little effect. We will discuss these issues
in more detail in an upcoming paper {\em VI} in the {\em star cluster
ecology} series (Portegies Zwart, McMillan \& Makino 2006+x \{with $x
\in \mathcal{I}; x\ge1$\}, in preparation).

\begin{figure}
\centering
   \includegraphics[height=10cm,angle=+90]{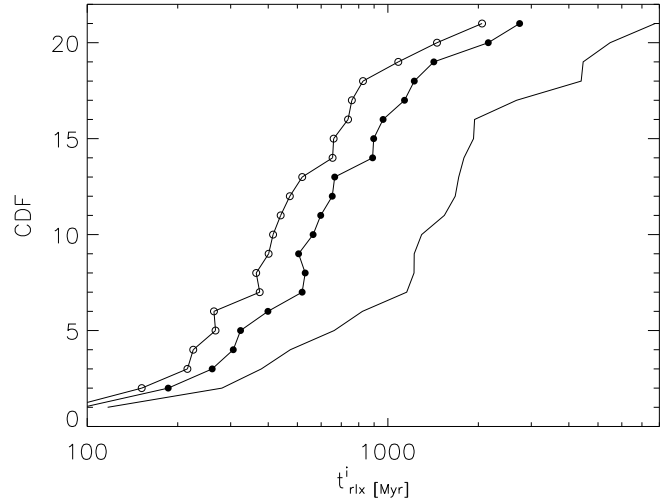}
   \caption{Cumulative distribution of the initial half-mass
   relaxation time for the observed sample of star clusters in the
   LMC. The open circles represent the distribution when the initial
   half-mass relaxation time was reconstructed using the simulations
   with 128k stars, the filled circles are reconstructed using the 12k
   simulations.  The solid curve without points (to the right) gives
   the observed distribution of present day relaxation times for the
   LMC clusters from data published by \cite{2003MNRAS.338...85M}.  We
   only used the data for those clusters that are younger than
   300\,Myr.
  \label{Fig:CDF_trlx_initial}}
\end{figure}

Based on the here presented simulations we reconstruct the initial
relaxation time for young ($\aplt 300$\,Myr) star clusters in the LMC
with the data from \cite{2003MNRAS.338...85M}.  In
Fig.\,\ref{Fig:CDF_trlx_initial} we present the cumulative
distribution of initial half-mass relaxation times for the observed
star clusters in the LMC. For comparison we overlaid the distribution
of the present day relaxation times for the observed clusters, which
ranges over more than two orders of magnitude. We conclude that the
initial relaxation time for the young ($\aplt 300$\,Myr) star clusters
in the large Magellanic cloud ranges from $\sim 200$\,Myr to $\sim
2$\,Gyr.

\section*{Acknowledgments}

I am grateful to Evgenii Gaburov, Mark Gieles, Alessia Gualandris and
Henny Lamers for many discussions.  This work was supported by NWO
(via grant \#630.000.001 and \#643.200.503), NOVA, the KNAW, the LKBF
and the following grands for the Taiwanese government under number
NSC095-2917-I-008-006 and NSC 95-2212-M-008-006.  The calculations for
this work were done on the MoDeStA computer in Amsterdam, which is
hosted by the SARA supercomputer center.

\end{document}